# Discovery of Hubble's Law: an Example of Type III Error

**Ari Belenkiy**

*Department of Statistics, Simon Fraser University, British Columbia, Canada*

## Abstract

Recently much attention has been paid to the discovery of Hubble's law – the linear relation between the rate of recession of the distant galaxies and distance to them. Though we now mention several names associated with this law instead of one, the motivation of each remains somewhat obscure. As it is turns out, two major contributors arrived at their discoveries from *erroneous reasoning*, thus making a case for a Type III error.

It appears that Lemaitre (1927) theoretically derived Hubble's Law due to his choice of the wrong scenario of the Universe's evolution. Hubble (1929) tested the linearity law not based on Lemaitre's non-static model, but rather a cumbersome extension of de Sitter's static theory proposed by Weyl (1923) and Silberstein (1924).

## Introduction

Though a conference was called at Flagstaff in 2012 to celebrate the centennial of V. Slipher's discovery of spectral shifts, little attention was given to the fact that in 1912 Slipher discovered only the blue-shift of the nearby Andromeda nebulae, which could hardly surprise (at least in hindsight) anyone familiar with gravitation theory. His truly surprising discovery of the spectral red-shifts, that later became a beacon for the discovery of the expansion of the Universe, occurred later, in 1913-1914. Thus, the proper date to celebrate is August 1, 1914, when the AAS met Slipher's report with a standing ovation.

## Slipher

While Slipher interpreted the Andromeda nebulae's blue-shift as a Doppler Effect caused by the nebulae's drift with respect to the Milky Way, as "we have no other interpretation for it,"[1] later, on discovering the red-shifts, he again suggested that the red-shifts might be due to the Milky Way's motion relative to the system of spiral nebulae. At that point he had mostly examined spirals in the northern part of the sky, so the fact that they were mostly red-shifted might just mean the Milky Way was moving in the opposite direction. But eventually spirals in the opposite part of the sky were examined and they too showed red-shifts, so this interpretation didn't hold up.

The first theoretician who paid close attention to Slipher's discovery, W. de Sitter (1917), denounced the red-shifts as indicators of the "spurious radial velocities."[2] This was an excellent insight in hindsight though based on incorrect reasoning – a Type III error – from his own theory that predicted a slowing of time at great distances. In 1923 H. Weyl concluded that "space objects have a natural tendency to scatter" but the reason for that "natural tendency" remained unknown to Weyl himself - as well as to A.S. Eddington who followed him – though de Sitter's theory again was their inspiration.[3]

**Friedman**

The first explanation capable of accounting for the supposed Doppler Effect conjectured by Slipher was suggested by A. Friedman (1922) – the Universe may expand since General Relativity (GR) equations admit dynamical solutions.[4] Alerted to this fact by P. Ehrenfest, Einstein missed the implications. Had the creator of General Relativity or W. de Sitter realized the implications, Einstein himself together with Friedman and Slipher could have been solid candidates for a Nobel Prize in Physics.

However, Einstein's famed intuition betrayed him this time. While being aware of Slipher's red-shift data for spirals, he did not pay them proper attention. Nobody at the time knew that spirals were distant galaxies. Einstein knew that the system of stars (within the Milky Way) showed no systematic motions toward or away from us, so he thought this indicated a static universe. His real problem was that he based his conclusion on the motion of objects at too small a distance scale.

This is an example of a Type II Error ("a miss"), where a wrong null-hypothesis of static Universe wasn't rejected despite strong evidence (the red-shifts) in favor of an alternative: a dynamic Universe.

When G. Lemaitre rediscovered the "Friedman equations" (and one of Friedman's expanding solutions) it was too late for the Nobel Prize – Friedman was dead (1925) and Einstein remembered very well who the first discoverer of the dynamic solutions was![5]

**Lemaitre**

Lemaitre could have qualified for the Nobel Prize on his own since he played the role Einstein missed – to connect Friedman's theory and Slipher's red-shifts. Lemaitre developed a test that links both: a linear relation between distances to galaxies *r* and their "radial velocities" *v*.[6]

It is interesting to trace how Lemaitre arrived at this idea. Noticing that the curvature radius *R(t)* of his metric is related to the red-shift *z* as

$$z = \dot{R}\, dt / R$$

and interpreting the time-interval *dt* as $r/c$, on the one side, and using the standard Doppler effect formula $z \approx v/c$, for $v << c$, on the other, Lemaitre was confronted with the equation

$$\dot{R} / R = v / r. \qquad (1)$$

This is a far cry from a linear ("Hubble") law – unless *R(t)* behaves exponentially over time! To find the exponential behavior Lemaitre had to commit a mistake choosing the wrong scenario of the Universe's evolution. Indeed, trying to "bridge" Einstein's (1917) and de Sitter's (1917) solutions, Lemaitre chose a solution of Friedman equations with a "logarithmic singularity" for time *t* as a function of radius *R* at a finite radius $R_E$, which has an infinitely long expansion $R \approx R_E + e^{\sqrt{\Lambda}(t-t_0)}$ from the finite ("Einsteinian") radius $R_E$ in the past ($t < t_0$) via the exponential growth $R \approx R_0 e^{\sqrt{\frac{\Lambda}{3}}(t-t_0)}$ in the future ($t > t_0$)) (see Fig. 1).[7] Such a nearly exponential behavior does not arise from any other solution – certainly, not from a Big Bang one!

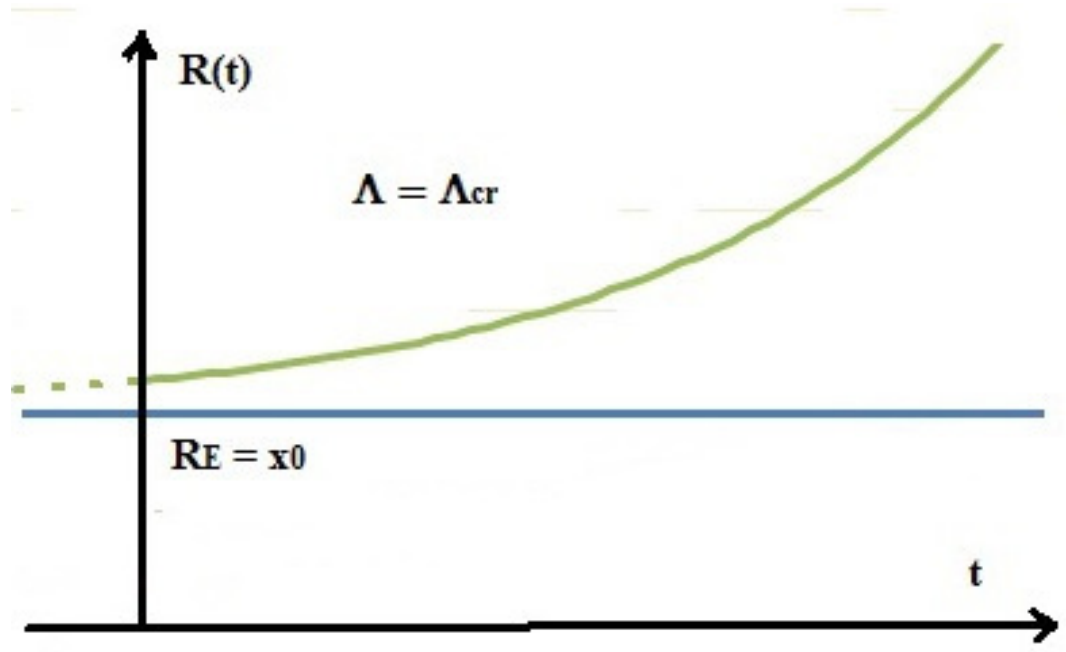

**Fig. 1**. A special solution of Friedman equations (Friedman 1922), with "logarithmic singularity" when viewing *t* as a function of *R*. The upper curve shows radius growing from RE nearly exponentially. The horizontal line stands for Einstein's static solution with constant curvature radius RE.

But now a remarkable feat comes. Though seeing a rather weak correlation between velocities, found by Slipher, and distances, provided by several Mount Wilson astronomers, for the set of 42 spiral galaxies, Lemaitre (1927) not only derived the linear relation from his theory, but made a reasonable estimate of the proportionality coefficient $v/r = H_1 = 575\, km \cdot s^{-1} \cdot Mpc^{-1}$. This is in fact the first empirically derived "Hubble constant".

That this derivation disappeared from the English translation with MNRAS (1931) recently caused much havoc, with suspicions raging to the point that E. Hubble personally edited Lemaitre's translation. These speculations have been rebuffed by the discovery of Lemaitre's letter (1931) to the editor of MNRAS where Lemaitre willingly agrees to drop his own estimate of H as new data became available:

*"I did not find advisable to reprint the provisional discussion of radial velocities which is clearly of no actual interest, and also the geometrical note, which could be replaced by a small bibliography of ancient and new papers on the subject."* [8]

This laid to rest the charges against Hubble's involvement, though one of the "new papers," mentioned by Lemaitre, was certainly due to Hubble (1929), who confirmed the linearity law by careful estimations of the distances to remote nebulae.[9] But how did Hubble embark on testing this law before becoming aware of Lemaitre's work?!

**Hubble**

Indeed, the drive behind Hubble's accomplishment was not based on Lemaitre's idea but an extension of de Sitter's theory by Weyl (1923) and L. Silberstein (1924). De Sitter's theory assumed constant radius *R*. But Weyl's computations of the body emitting light while moving on geodesics in de Sitter's universe led to a "nearly" linear law where the speed of light *c* appears instead of the rate of expansion $\dot{R}$:

$$c / R \approx v / r \quad , \tag{2}$$

where the right side carried additionally a quadratic term, which is negligible for $v \ll c$ and thus omitted.

Silberstein (1924) further added a seemingly "attractive" feature to Weyl's initial theory – a ± sign on the right side of (2) that might explain not only red-shifts but also blue-shifts observed for several nearby galaxies.[10] Most of the astronomers, however, denied that the available astronomical data supported the linear relation. Dubbed as the "Silberstein's effect," the latter fell in disrepute for some time.[11]

The things changed when Hubble (1926) mastered his method of computing the distances to the distant nebulae.[12] Since the left side of equation (2) was solidly constant, theoreticians, R.C. Tolman and H.P. Robertson, convinced Hubble to test the linear relation between two quantities in the right side.[13] Cautiously, trusting only galaxies distant from us at most by 2 Mpc, Hubble (1929) indeed produced two famous straight lines where one line was based on an equal individual weighting of 46 galaxies, while the other line was obtained by first combining the galaxies into nine groups, averaging *v* and *r* for each group, and then calculating the fit by giving equal weight to each group. The slope of the first line, $v/r = H_2 = 525\, km \cdot s^{-1} \cdot Mpc^{-1}$,[14] soon became known as the "Hubble constant."

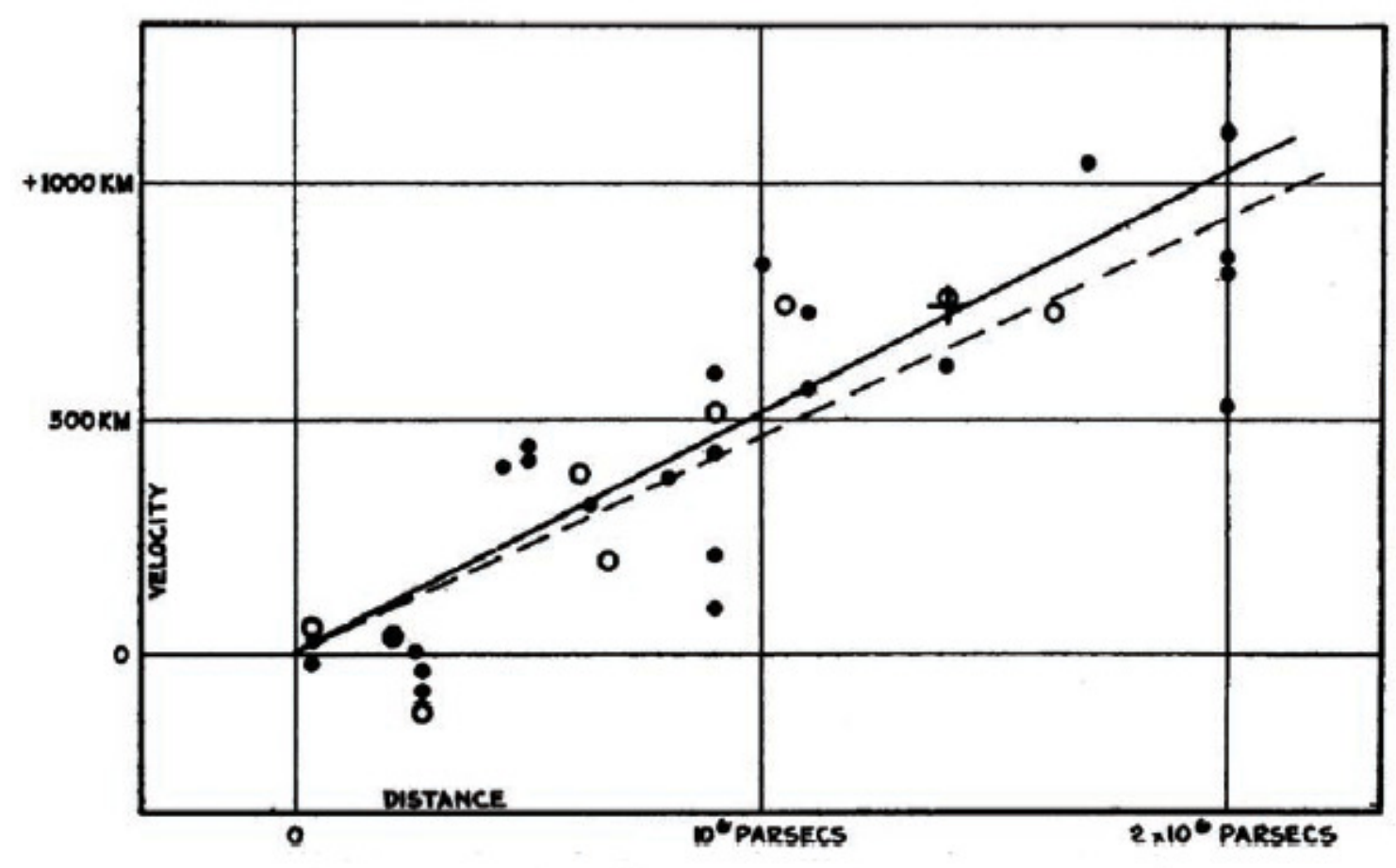

**Figure 2**. The velocity-distance linear relation for spiral galaxies found by E. Hubble in 1929. The slopes of the lines (derived in two ways for the same set of data) provide the estimates for the *proportionally coefficient* between velocity and distance that became later known as the "Hubble constant."

This result led to a remarkable finale. In January 1930, Eddington and de Sitter publicly rejected the very inspiration for Hubble's investigation – de Sitter's theory as inadequate to explain the linear law![15] This opened the door for a new paradigm. Shortly thereafter Lemaitre reminded Eddington of his 1927 paper, which was immediately hailed by the latter as a great discovery.

**Type III Error cases for Lemaitre and Hubble**

Despite their seminal discoveries made partially because of mistaken assumptions, neither Lemaitre immediately nor Hubble ever renounced their mistaken beliefs.

In his 1930 letter to Eddington, Lemaitre writes:

"I consider a Universe of curvature constant in space but increasing with time and I emphasize the existence of a solution in which the motion of the nebulae is always a receding one from time minus infinity to plus infinity."[16]

Thus, as late as 1930 Lemaitre still adhered to the "limiting case" scenario. He abandoned it the following year -- because Eddington pointed out that "such logarithmic singularities have no physical significance" – but only in favor of another model (the "monotone world of the second kind," in Friedman's terminology), which starts from non-zero initial ("Einsteinian") radius $R_{min}$ with nearly exponential behavior of *R(t)* for t > 0 (see M2 in Fig. 2). [17] Only later that year did Lemaitre begin to consider the model with initial singularity, a prototype of the Big Bang model (the "monotone world of the first kind," in Friedman's terminology).[18] This model is far from being exponential, especially near the flex point $t_f$ (see M1 in Fig. 2).

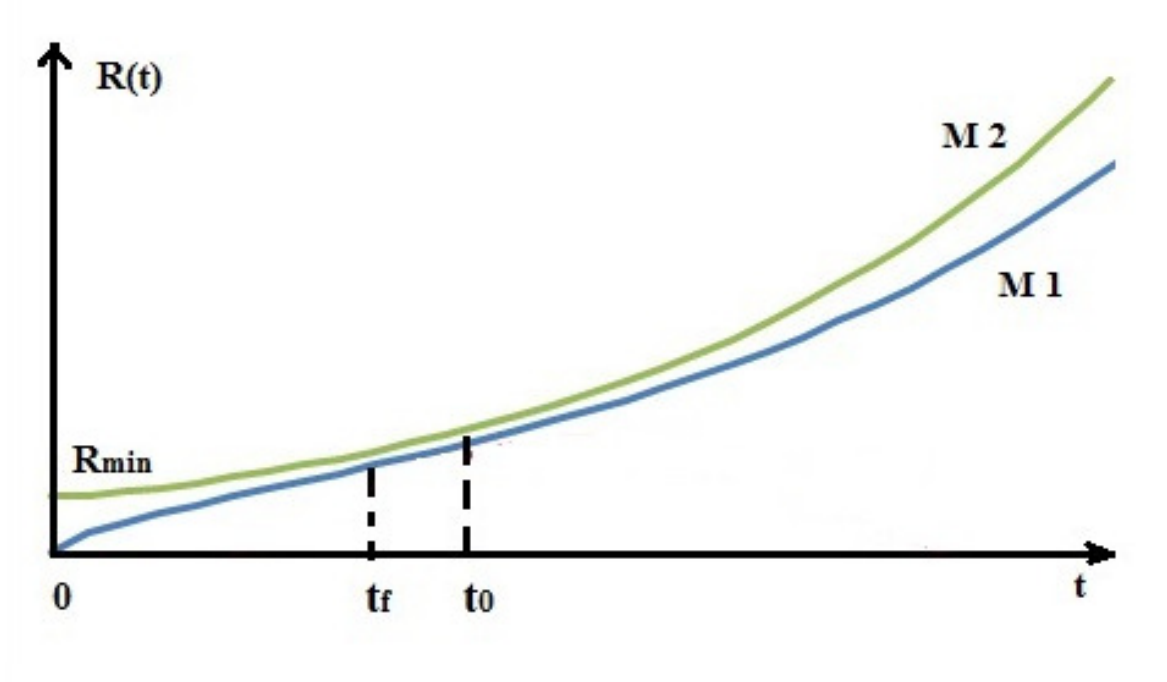

**Fig. 3**. Two possible major scenarios of the Universe's evolution according to Friedman (1922). The M1 world shows expansion from singularity at t=0 with an inflexion point $t_f$ that signifies existence of two stages of evolution: deceleration and acceleration. The M2 world shows expansion from the non-zero initial radius $R_{min}$ to infinity. Point $t_0$ is the current stage of the Universe.

As we know Hubble never accepted the interpretation of his discovery as "expansion of the universe," as

*"it is difficult to believe that the velocities are real; that all matter is actually scattering away from our region of space. It is easier to suppose that the light waves are lengthened and the lines of the spectra are shifted to the red, as though the objects were receding, by some property of space or by forces acting on the light during its journey to the Earth."*[19]

Of course, he interpreted his discovery as an opportunity to evaluate "constant" radius *R* from the equation (2). Ironically, the rate of expansion $\dot{R}$ and c are indeed very close in modern time, making this approach seemingly credible.

Indeed, following Hubble in another of the “new papers” mentioned by Lemaitre, de Sitter (1930) derived proportionality parameter $v/r = H_3 = 425\,km\cdot s^{-1}\cdot Mpc^{-1}$ and found from (2) $R = 2\cdot 10^9$ light years.[20] On the other side, the initial Big Bang theory developed by Lemaitre (1934), led to approximately the same present radius, $R(t) = 1.7\cdot 10^9$ light years,[21] which was in odds with the geological age of the Earth. Since both estimates depended on Hubble’s estimate of the distances to the spiral nebulae, Hubble, together with the part of the astronomical community, could remain “reasonably” skeptical toward the idea of an expanding universe. Indeed though recognizing that the “expanding universe, with its momentary dimensions as previously described, is the latest widely accepted development in cosmology,” Hubble still believed that

*“the 200-inch telescope will definitely answer the question of the interpretation of red-shifts, whether or not they represent actual motions, and if they do represent motions –if the universe is expanding – the 200-inch may indicate the particular type of expansion.”*[22]

- thus making a case for Type I Error (a “false alarm”), incorrectly rejecting the dominant null-hypothesis of expanding universe in favor of an alternative, the supposed “tiredness of light coming from afar.”

**Summary**

The discovery of non-static solutions to Einstein's GR equations was first made by A. Friedman (1922) and later by G. Lemaitre (1927). Observational support for these solutions were provided by the red-shift measurements of V.M. Slipher in 1910s and the distance measurements of E. Hubble in 1920s. Lemaitre (1927) was the first to realize that the linear relationship between the distance and velocity of spiral nebulae can be a test for the non-static solution of Friedman/Lemaitre. The linear relationship between velocity and distance was confirmed by Hubble first in early 1929 and later, in association with M. Humason, in 1931. Hubble’s confirmation was not associated with non-static solutions until early in 1930 when A. Eddington and W. de Sitter became aware of Lemaitre's 1927 paper. The father of GR, Einstein, conceded this idea only in early 1931, advancing two new hypotheses, which however are out of the scope this paper.[23]

Thus said, errors can still lead to progress. The errors made on the way to the discovery of the expanding universe by the pioneers of the modern cosmology did not prevent the cosmological community from "getting it right" eventually. In fact, they were crucial steps in our progress toward understanding the cosmic expansion. Science progresses by correcting errors, not by avoiding them altogether.

**Acknowledgements**

The author thanks Cormac O'Raifeartaigh (Waterford Institute of Technology, Ireland), Todd Timberlake (Berry College) and Michael Way (NASA) for helpful discussion and valuable comments.

---

[1] V. M. Slipher, 1913, "The radial velocity of the Andromeda nebula," *LowOB* **2**, 56.

[2] W. de Sitter, 1917, "Einstein's theory of gravitation and its astronomical consequences. Third paper," *MNRAS* **78**, 3-28.

[3] H. Weyl, 1923, *Raum, Zeit, Materie.* Berlin, Springer, fifth edition, 322-323. Also: H. Weyl, 1923, "Zur algemeinen Relativifatstheorie," *Phys. ZS*. **24**, 230. Also: A. S. Eddington, 1923, *The Mathematical Theory of Relativity*. Cambridge, University Press.

[4] A. Friedman, 1922, "Uber die Krummung des Raumes," *Zeitschrift f. Physik* **10**, 377-387, or in English translation: A. Friedman, "On the curvature of space." *General Relativity and Gravitation* **31** (12), 1991–2000 (1999).

[5] A. Belenkiy, 2012, "Alexander Friedmann and the Origins of Modern Cosmology," *Physics Today* **65**, 38.

[6] G. Lemaitre, 1927, "Un univers homogene de masse constante et de rayon croissant, rendant compte de la vitesse radiale des nebuleuses extra-galactiques," Ann. de la Soc. Scientif. de BruxelleS **47***,* 49.

[7] A. Belenkiy, 2013, "The waters I am entering no one yet has crossed: Alexander Friedman and the origins of modern cosmology." In: "Origins of the Expanding Universe: 1912-1932", M. J. Way & D. Hunter, eds., ASP Conf. Ser. **481**, 81-98.

[8] M. Livio, 2011, "Lost in Translation: Mystery of the Missing Text Solved," *Nature* **479**, 171.

[9] E. Hubble, 1929, "A Relation between Distance and Radial Velocity among Extra-Galactic Nebulae," *PNAS* **15**, 168-173.

[10] L. Silberstein, 1924, "The Radial Velocities of Globular Clusters and de Sitter's Cosmology," *Nature* **113**, 2836, 350.

[11] See H. Kragh, 1999, *Cosmology and Controversy: the Historical Development of Two Theories of the Universe*, Princeton University Press, p. 15.

[12] E. Hubble, 1926, ""Extragalactic nebulae," *ApJ* **64**, 321-369 (1926).

[13] R. C. Tolman, 1929, "On the astronomical implications of the de Sitter's line element." *ApJ* **69**, 245-274. See also H. Nussbaumer & L. Bieri, 2009. *Discovering the Expanding Universe* (Cambridge University Press), p. 118.

[14] E. Hubble, 1929, "A Relation between Distance and Radial Velocity among Extra-Galactic Nebulae," *PNAS* **15**, 168.

[15] W. de Sitter, 1930, "Proceeding of the R.A.S.," *The Observatory* **53**, 37-39. See also H. Nussbaumer & L. Bieri, 2009. *Discovering the Expanding Universe*, p. 121.

[16] H. Nussbaumer & L. Bieri, 2009, *Discovering the Expanding Universe*, p. 122.

[17] G. Lemaitre, 1931, "The Expanding Universe," *MNRAS* **91**, 490-501.

[18] G. Lemaitre, 1931, "The Beginning of the World from the Point of View of Quantum Theory," *Nature* **127**, 3210, 706.

[19] E. Hubble, 1929, "A clue to the structure of the Universe," *ASP Leaflets* **23**, 93-96. See also H. Nussbaumer & L. Bieri, 2009, *Discovering the Expanding Universe*, pp. 119-120.

[20] W. de Sitter, 1930, "On the magnitudes, diameters and distances of the extragalactic nebulae and their apparent radial velocities (Errata: 5 V, 230)," *BAN* **5**, 157-171.

[21] G. Lemaitre, 1934. "Evolution of the Expanding Universe," *PNAS* **20**, 12-17.

[22] E. Hubble, 1934, "The Realm of the Nebulae," *The Scientific Monthly* 39 (3), 193-202.

[23] See, however, forthcoming papers by H. Nussbaumer and C. O'Raifeartaigh at EJP-H.